\documentclass[11pt]{article}
\usepackage{moriond,epsfig}

\def\be{\begin{equation}}
\def\ee{\end{equation}}
\def\bea{\begin{eqnarray}}
\def\eea{\end{eqnarray}}

\begin{document}
\vspace*{4cm}
\title{SMALL-X PHYSICS NEAR THE SATURATION REGIME}

\author{A.H. MUELLER and A.I. SHOSHI}

\address{Department of Physics, Columbia University, New York, NY
  10027, USA}

\maketitle\abstracts{ We consider the $T$-matrix near the unitarity
  limit and the energy dependence of the saturation momentum. We
  discuss the solution to the Kovchegov equation, or equivalently, to
  the BFKL evolution in the presence of a single saturation boundary.
  We include some of the correlations missed in the Kovchegov equation
  by solving the BFKL equation in the presence of two boundaries. The
  $T$-matrix now turns out to be frame-independent, which was not the
  case for the solution in the case of a single boundary, and it
  doesn't show the scaling behavior of the solution to the Kovchegov
  equation. We find for the saturation momentum an energy dependence
  which differs from the one following from the Kovchegov equation.}

\section{Introduction}

The growth of cross sections with increasing energy, or parton
densities with decreasing Bjorken-$x$, at a fixed hard scale is given
by the BFKL~\cite{Kuraev:fs+X} evolution when parton occupation
numbers are not too large or the $T$-matrix not to close to the
unitarity limit.  At or near the unitarity limit nonlinear parton
evolution becomes important which is not described by the linear BFKL
equation. An extension of the BFKL equation that includes also
nonlinear evolution is the Balitsky equation~\cite{Balitsky:1995ub+X}
or the Jalilian-Marian, Iancu, McLerran, Leonidov and Kovner
(JIMWLK)~\cite{Jalilian-Marian:1997jx+X} equation.  The Balitsky and
JIMWLK equations are coupled equations involving higher and higher
correlations and as such are very difficult to deal with analytically.
A better understanding of the Balitsky and JIMWLK evolution
may emerge from numerical calculations such
as~\cite{Rummukainen:2003ns}.
 
Kovchegov~\cite{Kovchegov:1999yj+X} has suggested a simpler equation
than the Balitsky or JIMWLK equations to deal with scattering at or
near the unitarity limit. The Kovchegov equation can be viewed as a
mean field version of the Balitsky or JIMWLK equation in which higher
correlations are neglected. While incomplete, as is any mean field
like approximation, the Kovchegov equation is likely the best equation
one can write down in terms of functions which has built in correct
unitarity limits for high energy scattering. Many interesting limits
of the Kovchegov equation have been understood by analytical methods.
In this work we focus on the transition region from weak to the
saturation regime in which Munier and
Peschanksi~\cite{Munier:2003vc+X} have determined the form of the
$T$-matrix and the rapidity dependence of the saturation momentum by
solving the Kovchegov equation analytically. The same results have
been obtained even before by the authors of Ref.~\cite{Mueller:2002zm}
by solving the BFKL equation in the presence of a single saturation
boundary which effectively approximates the unitarity limit guaranteed
in the Kovchegov equation.  In the next Section we derive the
Kovchegov equation and study its limitations due to the missed higher
correlations.

We attempt to go beyond the mean field like approximation in the
Kovchegov equation by solving the BFKL equation in the presence of two
boundaries~\cite{Mueller:2004se}. We find in the vicinity of the
saturation regime an expression for the $T$-matrix which is
frame-independent and doesn't show a scaling behavior in contrast to
the solution to the Kovchegov equation. The energy dependence of the
saturation momentum is also different from the one coming
from the Kovchegov equation.

\section{The Kovchegov equation}
Consider the high-energy scattering of a color dipole on a target
(another dipole, hadron, or nucleus) at relative rapidity $Y$ in a
frame where the dipole is an elementary quark-antiquark pair and the
target a highly evolved system. Now we wish to know how the elastic
scattering amplitude $S(\underline{x}_0,\underline{x}_1,Y)$ changes
when the rapidity $Y$ is increased by a small amount $dY$ (${\underline x}_0$ and ${\underline x}_1$ are the transverse
coordinates of the quark and antiquark of the dipole). The change of
$S(\underline{x}_0,\underline{x}_1,Y)$ with rapidity is determined by  
the Balitsky or JIMWLK equation
\be 
\frac{\partial}{\partial Y} S(\underline{x}_{01},Y) =
           \frac{\alpha N_c}{2 \pi^2} \int d^2{\underline x}_2
           \frac{\underline{x}^2_{01}}{\underline{x}^2_{02} 
           \underline{x}^2_{12}} \left [
           S^{(2)}(\underline{x}_{02},\underline{x}_{12},Y) -
           S(\underline{x}_{01},Y) \right ] \ .
\label{eq_B_JIMWLK}
\ee
This equation can be interpreted as follows: When the increase $dY$ is
viewed as increasing the rapidity of the dipole then the
probability for the dipole to emit a gluon (transverse coordinate
$\underline{x}_2$) increases. In the large $N_c$ limit this
quark-antiquark-gluon state can be viewed as a system of two dipoles.
The scattering of this two dipole state on the target is given by
$S^{(2)}(\underline{x}_{02},\underline{x}_{12},Y)$ while the
subtracted $S(\underline{x}_{01},Y)$ is the virtual contribution
necessary to normalize the
wavefunction~\cite{Mueller:1994rr+X}. The later gives the
scattering of a single dipole on the target because the gluon is not
in the wavefunction of the dipole at the time of the scattering. The
weight in Eq.~\ref{eq_B_JIMWLK} gives the probability for the
emission of two dipoles of sizes $\underline{x}_{02}$ and
$\underline{x}_{12}$ from the initial dipole of size
$\underline{x}_{01}$.

It is difficult to use the Balitsky-JIMWLK equation because of the
unknown $S^{(2)}(\underline{x}_{02},\underline{x}_{12},Y)$. The
assumption that the scattering of the two dipole state on the target
factorizes
\be
S^{(2)}(\underline{x}_{02},\underline{x}_{12},Y) =
  S(\underline{x}_{02},Y) S(\underline{x}_{12},Y) \ ,
\label{fac}
\ee
which is a sort of a mean field approximation for the gluonic fields
in the target, leads to the Kovchegov
equation~\cite{Kovchegov:1999yj+X}
\be 
\frac{\partial}{\partial Y} S(\underline{x}_{01},Y) =
           \frac{\alpha N_c}{2 \pi^2} \int d^2{\underline x}_2
           \frac{\underline{x}^2_{01}}{\underline{x}^2_{02} 
           \underline{x}^2_{12}} \left [
           S(\underline{x}_{02},Y) S(\underline{x}_{12},Y) -
           S(\underline{x}_{01},Y) \right ] \ .
\label{Eq_Kovchegov}
\ee
%
 
One exact result of the Kovchegov equation
for the $S$-matrix deep in the saturation
region is the Levin-Tuchin
formula~\cite{Levin:1999mw} 
\be
S(\rho,b,Y) \sim e^{-c (\rho-\rho_s)^2}
\label{eq_S_c}
\ee
with the constant $c=-C_F(1-\lambda_0)/(N_c 2 \chi(\lambda_0))$.
Recently, the authors of Ref.~\cite{Iancu:2003zr} have claimed that
$S$ deep in the saturation regime has the form given by
Eq.~\ref{eq_S_c} but with a constant at least a factor of $2$ smaller
than the $c$ which follows from the Kovchegov equation. The cause for
this discrepancy is the lack of fluctuations in the Kovchegov
equation. 

Now let us focus on the region close to the saturation regime.
Consider the scattering of a dipole of size $x$ on a dipole of size
$x'$ at relative rapidity $Y$ and impact parameter $b$.  Near the
saturation boundary the Kovchegov equation  or the BFKL
equation in the presence of an absorbtive saturation boundary give for the
$T$-matrix in laboratory frame
\be
T(x,x',Y;b) = T_0(b,x)\ \left[Q_s^2(Y)\,x'^2\right]^{1-\lambda_0}\
            \ln\frac{1}{Q^2_s(Y)\,x'^2}\ \exp\!\left[-\frac{\pi
                \ln^2(1/Q^2_s(Y)\,x'^2)}{4 \alpha N_c
                \chi''(\lambda_0)Y}\right]  
\label{eq_T_fQ}
\ee
and for the rapidity dependence of the saturation momentum 
\be
Q^2_s(x,Y;b) = Q^2_0(x,b)\ \frac{1}{x^2}\ \frac{\exp\!\left[\frac{2\alpha
      N_c}{\pi}\frac{\chi(\lambda_0)}{1-\lambda_0}Y\right]}
     {\left[\alpha Y\right]^{\frac{3}{2
           (1-\lambda_0)}}} 
\label{eq_Q_s}
\ee
where $\lambda_0=0.372$. Near the saturation boundary, i.e., for the
size of the probe $x'^2$ close to $1/Q^2_s(Y)$, the scattering
amplitude in (\ref{eq_T_fQ}) shows a scaling behavior since it depends
on $Q_s^2(Y)\,x'^2$; lines with constant $Q_s^2(Y)\,x'^2$ are lines of
constant scattering amplitude.

Recently~\cite{Mueller:2004se} we have shown that fluctuations are
important in evolution also in the region near the saturation boundary
(``scaling region''). We have found that the completeness relation
\be
n(x,x',Y) = \int\!\!\frac{d^2r_2}{2 \pi r_2^2}\ n(x,r_2,Y/2)\ 
n(r_2,x',Y/2) \ ,
\label{eq_cr}
\ee 
is not satisfied for the dipole number density $n$ which is obtained
by using the BFKL evolution in the presence of a single saturation
boundary, as in the case of the $T$-matrix in Eq.~\ref{eq_cr}. This
mismatch is equivalent to the statement that the result for the
$T$-matrix in different frames is different. The reason for this
mismatch is the mean field like treatment of fluctuations in the
Kovchegov equation: On the left hand side of Eq.~\ref{eq_cr} the
evolution from the initial point ($x, y=0$) to the final point ($x',
y=Y$) occurs in one step while on the right hand side of
Eq.~\ref{eq_cr} the evolution can be viewed as proceeding in two
steps, form ($x, y=0$) to ($r_2, Y/2$) to ($x', Y$), and at the
intermediate stage of evolution the fluctuations are not properly
taken into account.  

\begin{figure}[ht]
\setlength{\unitlength}{1.cm}
\begin{center}
\epsfig{file=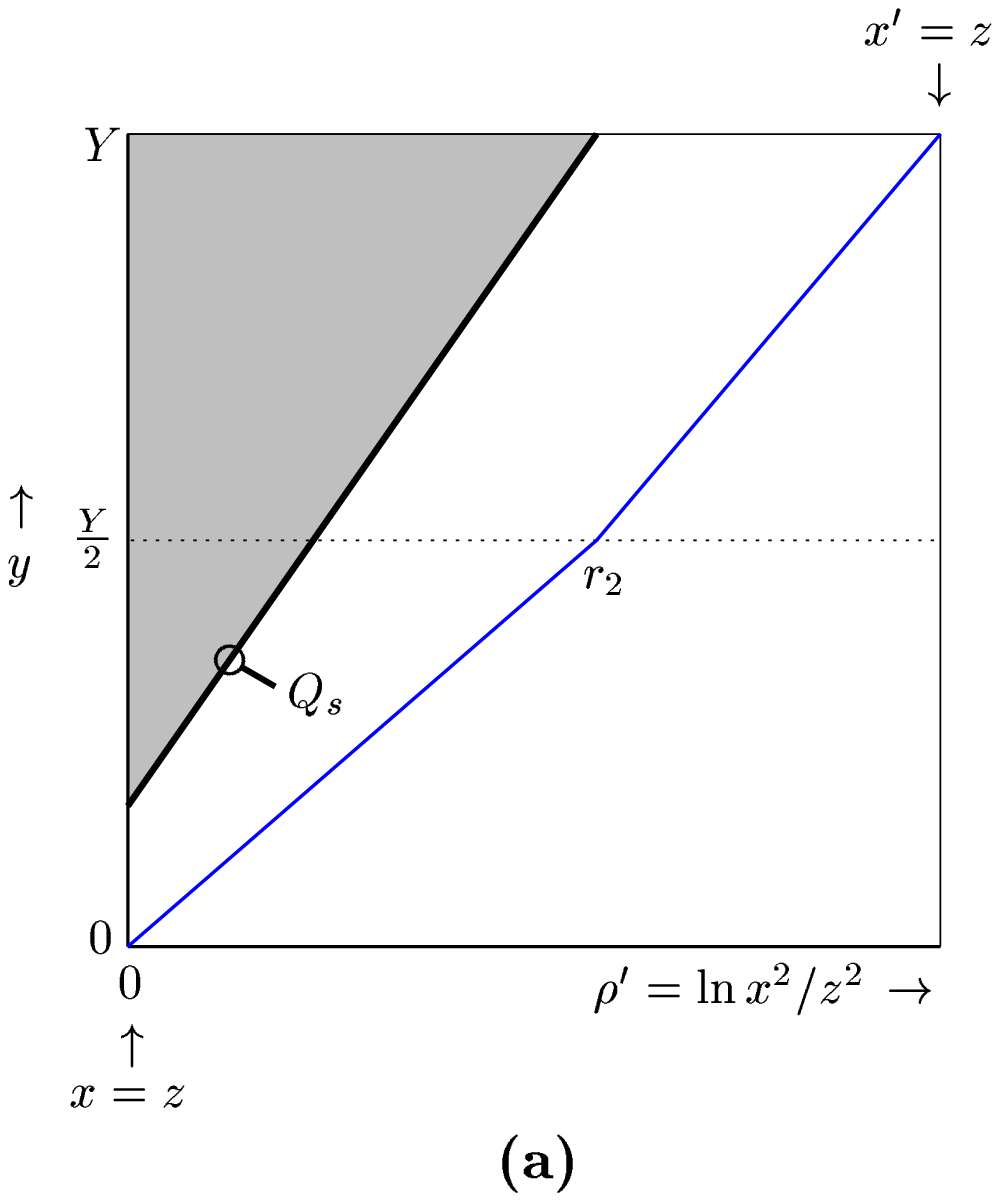, width=5cm} \hfill
\epsfig{file=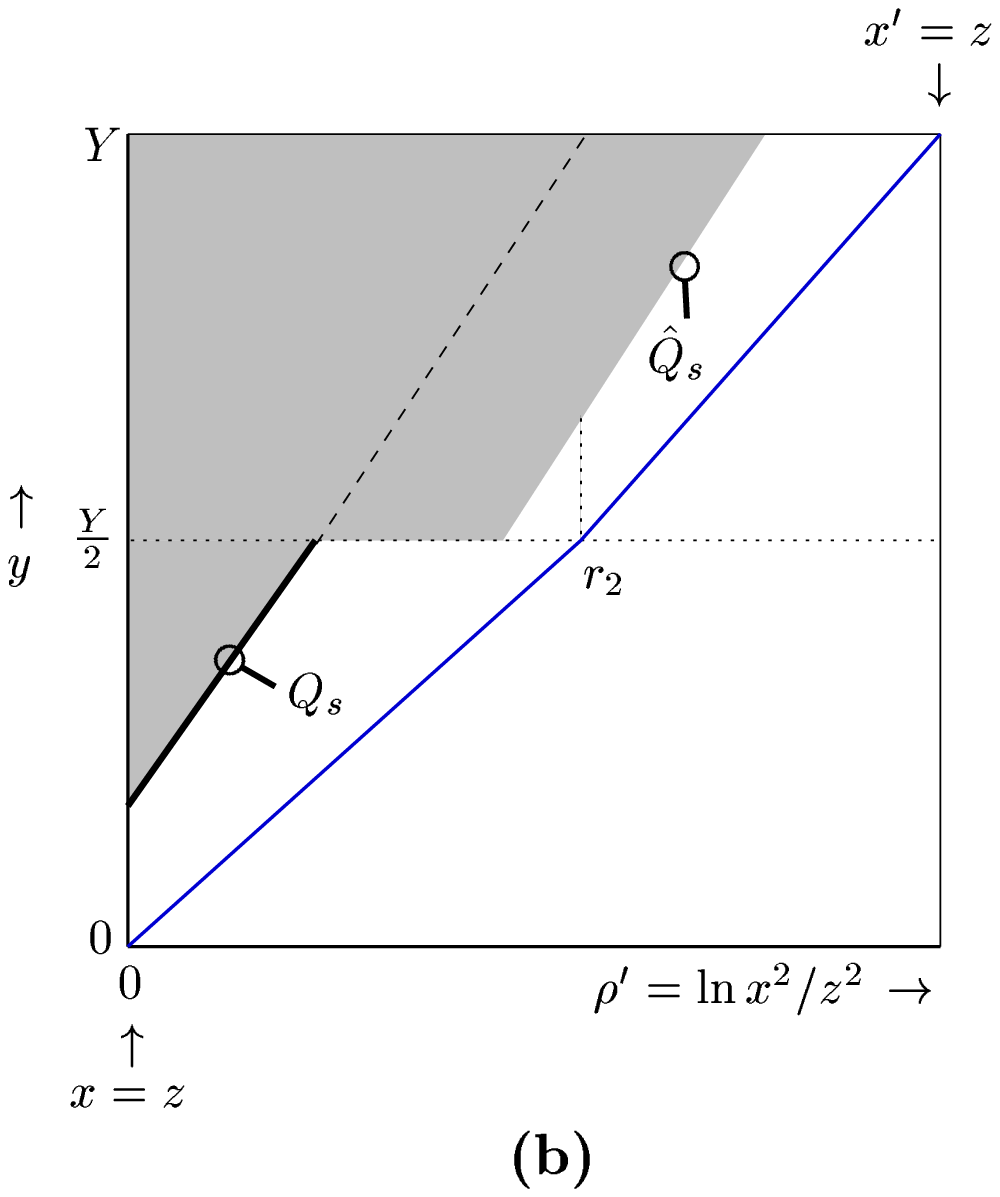, width=5cm} \hfill
\epsfig{file=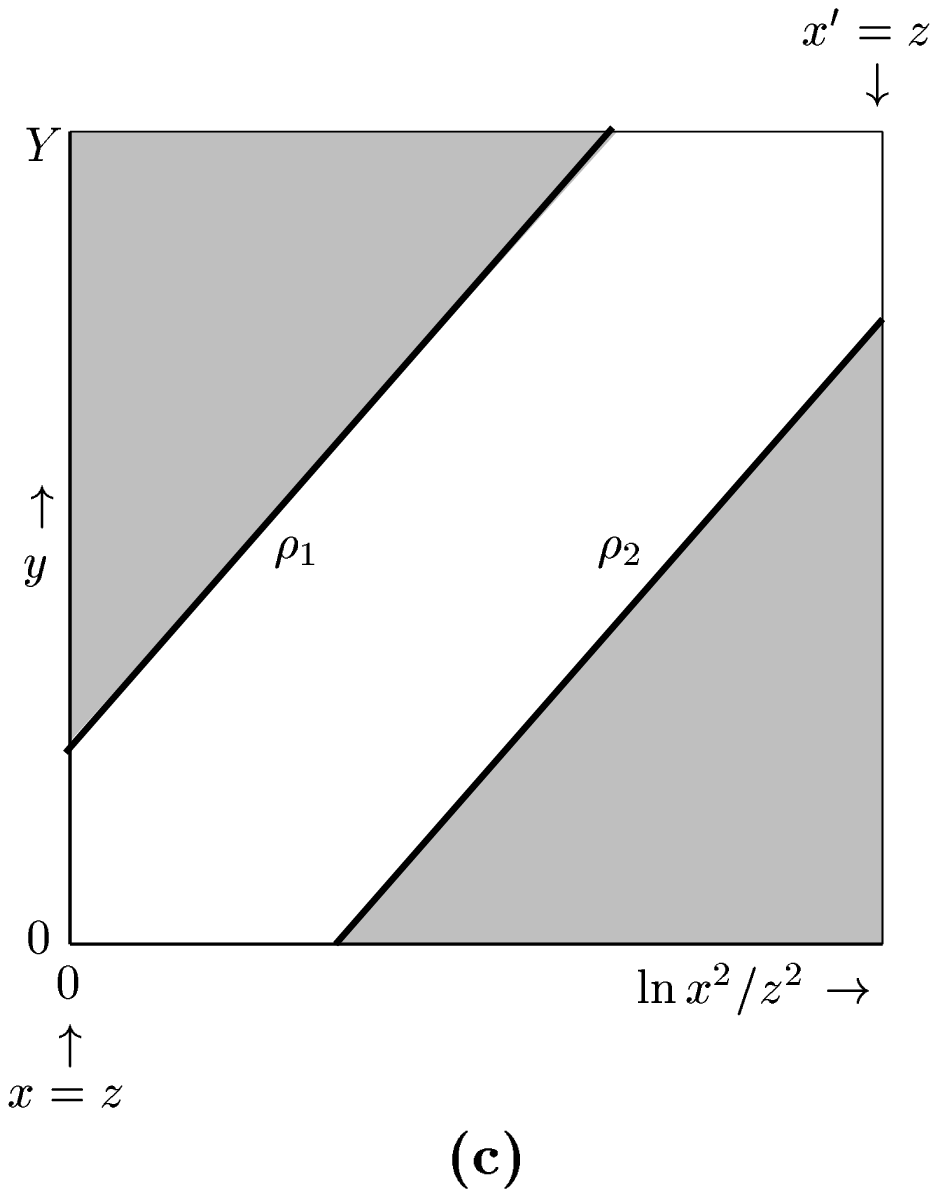, width=4.7cm}
\end{center}
\caption{BFKL evolution in the presence of shaded saturation regions
  in the $Y-\ln(x^2/z^2)$ plane with $z$ generic: (a) evolution in the
  presence of a single saturation boundary, (b) the second evolution
  step from $Y/2 \to Y$ has its own saturation boundary, (c)
  evolution in the presence of two boundaries.}
\label{Fig_1}
\end{figure}

In terms of BFKL evolution in the presence of a saturation boundary
the reason why the completeness relation is not satisfied goes as
follows: on the lhs of Eq.~\ref{eq_cr} the entire evolution occurs in
the presence of a fixed boundary $Q_s$ determined by $x$ and $y$ (Fig.~\ref{Fig_1}a) while on the rhs of Eq.~\ref{eq_cr} the second part
of evolution occurs in the presence of a different boundary
$\hat{Q}_s$ determined by $r_2$ and $y$ (Fig.~\ref{Fig_1}b). Since
the dipole number density is obtained in the mean field approximation,
one expects that the completeness relation should be fulfilled in an
approximative way. Indeed, if one requires for the evolution on both
sides of Eq.~\ref{eq_cr} to happen in the presence the same saturation
boundary (Fig.~\ref{Fig_1}a), i.e., if one extends the diffusion
region for the second part of the evolution on the rhs of
Eq.~\ref{eq_cr} from $\hat{Q}_s$ to $Q_s$, then the lhs and rhs of
Eq.~\ref{eq_cr} give the same result. However, because of the extended
diffusion region, there are now paths of evolution from ($r_2, Y/2$)
to ($x', Y$) which manifestly violate unitarity ($T\gg1$). See
for more details~\cite{Mueller:2004se}. 
\section{BFKL evolution in the presence of two saturation boundaries}\label{sec:ssb}
Now we introduce a second absorptive barrier at $\rho_2(y)$, as shown
in Fig.~\ref{Fig_1}c, such that, all unitarity violating evolution
between the initial point ($x$, $0$) and the final point
($x'$, $Y$) is eliminated when the evolution is viewed as
proceeding in two steps~\cite{Mueller:2004se}. The boundary
$\rho_1(y)$ corresponds to saturation in the wavefunction of the
evolved dipole $x$ scattering on an elementary dipole $x'$ (forward
evolution) and $\rho_2(y)$ corresponds to saturation in the
wavefunction of the evolved dipole $x'$ scattering on the elementary
dipole $x$ (backward evolution). The introduction of the second
boundary $\rho_2(y)$ and the symmetry it brings with has another
benefit: The $T$-matrix is now frame-independent which was not the
case for the single boundary case.

Near the saturation regime the BFKL evolution in the presence of two
absorbtive boundaries~\cite{Mueller:2004se} gives a $T$-matrix which
depends on $\ln(Q_s(Y)\,x')/[\alpha Y/(\Delta\rho)^3]$ with
$\Delta\rho \approx \frac{2}{1-\lambda_0}\ \ln(1/\alpha)$. Thus, the
scaling behavior of the solution to the Kovchegov equation is now
lost. The rapidity dependence of the saturation momentum for $\alpha Y
\gg (\rho_2-\rho_1)^2/(\pi N_c \chi''(\lambda_d))$ becomes
\be
Q^2_s(x,Y;b) = Q_0^2(x,b)\,\frac{1}{x^2}\, \exp\left[\frac{2\alpha
             N_c}{\pi}\frac{\chi(\hat{\lambda}_d)}{1-\hat{\lambda}_d} 
            \left(1-{\frac{\pi^2}{2
            \Delta\rho^2}\frac{\chi''(\hat{\lambda}_d)}{\chi(\hat{\lambda}_d)}(1+\frac{c_s}{\Delta\rho})}\right) Y\right]
\label{eq_BFKL_DW_Q_S}
\ee
with $\hat{\lambda}_d = \lambda_0+\pi^2/[2 (\Delta\rho)^2
(1-\lambda_0)]$ and is different as compared to the result in Eq.~\ref{eq_Q_s}.

\section*{Acknowledgments}
A.~Sh. is supported by the Deutsche Forschungsgemeinschaft under
contract Sh 92/1-1.

\end{document}